\documentstyle[graphicx,latexsym,times]{mn}

\title[Luminosity
functions of galaxies in the Coma cluster]{\emph{U}, \emph{B} and \emph{r} band luminosity functions
of galaxies in the Coma cluster\thanks{Based on observations made with the Isaac Newton Telescope, operated on the island of La Palma by the Isaac Newton Group in the Spanish Observatorio del Roque de los Muchachos of the Instituto de Astrofisica de Canarias}}
\author[Marco Beijersbergen et
al.]{Marco Beijersbergen,$^{1}$\thanks{E-mail: beijersb@astro.rug.nl} Henk
 Hoekstra,$^{2,3}$ Pieter G.~van~
Dokkum,$^{4}$\thanks{Hubble Fellow} \newauthor and Thijs (J.M.)~van~der~Hulst$^{1}$
\\
$^1$Kapteyn Astronomical Institute, P.O. Box 800, 9700 AV
       Groningen, The Netherlands\\ 
$^2$CITA, 60 St. George Street, Toronto, M5S
3H8, Canada\\
$^3$Department of Astronomy, University of Toronto, 60 St. George Street,
Toronto, M5S 3H8, Canada\\ 
$^4$California Institute of Technology,
Mail Stop 105-24, 1200 E California Blvd Pasadena, CA 91125, USA\\
}

\date{Version \today}

\pubyear{2001}

\begin{document}

\maketitle

\begin{abstract}
We present a deep multi-colour CCD mosaic of the Coma cluster (Abell 1656), covering 5.2 deg$^2$ in the $B$ and $r$ bands, and 1.3 deg$^2$ in the $U$ band. This large, homogeneous data set provides a valuable low redshift comparison sample for studies of galaxies in distant clusters. In this paper we present our survey, and study the dependence of the galaxy luminosity function (LF) on passband and radial distance from the cluster centre. The
$U$, $B$ and $r$ band LFs of the complete sample cannot be represented
by single Schechter functions. For the central area, $r<245
~h_{100}^{-1}$ kpc, we find best-fitting Schechter parameters of $M^{*}_{\rm U}=-18.60^{+0.13}_{-0.18}$ and
$\alpha_{\rm U}=-1.32^{+0.018}_{-0.028}$, $M^{*}_{\rm B}=-19.79^{+0.18}_{-0.17}$ and
$\alpha_{\rm B}=-1.37^{+0.024}_{-0.016}$ and $M^{*}_{\rm
r}=-20.87^{+0.12}_{-0.17}$ and $\alpha_{\rm
r}=-1.16^{+0.012}_{-0.019}$. The LF becomes steeper at larger radial
distance from the cluster centre. The effect is most pronounced in the $U$ band. This result is consistent
with the presence of a star forming dwarf population at large distance
from the cluster centre, which may be in the process of being accreted
by the cluster. The shapes of the LFs of the NGC 4839 group support a scenario
in which the group has already passed through the centre.
\end{abstract}


\section{Introduction}

Clusters of galaxies are important laboratories for studies of galaxy evolution. The galaxy population in clusters is very different from the
population in the field, suggesting that galaxy formation and
evolution are a strong function of the environment. 

The galaxy luminosity function (LF) should be an excellent tracer of environmental effects. Knowledge of the shape of the LF (in different bands) is
a powerful tool for studies of galaxy evolution. Specifically, one can look for
correlations between the shape of the general LF and environmental or cluster
properties. In general the LF drops steeply at bright magnitudes and
rises gradually at fainter magnitudes as described by a Schechter function \cite{schechter}. Sometimes, however, features such as bumps
and a steeply rising faint part are found, which cannot be adequately
fitted by a single Schechter function. Galaxies in the cluster core region are expected to have a
different merger history than the galaxies populating the cluster
outskirts where it blends into the field. This should be reflected in
differently shaped LFs for dense and less dense regions within a
cluster. L\'{o}pez-Cruz et al. \shortcite{lopez-cruz_etal} propose that the flat faint end slopes found in rich clusters result from the disruption of dwarf galaxies. Biviano et al. \shortcite{biviano_etal} report a dip in
the bright part of the general LF for rich clusters which is not seen 
in LFs of poor clusters or in the field. Andreon \shortcite{andreon} verified the invariance of the shape
of the bright part of the type-dependent LF in a large range of environments from
the field to the cores of clusters several orders of magnitude
denser. The determination
of the exact shapes of LFs is difficult as the faint ends suffer from
uncertainties in the contamination by field galaxies. 

The Coma cluster
($z=0.023$, richness class 2) is the
richest of the nearby clusters and ideal to study environmental
effects (for a detailed
overview of research on the Coma cluster see Mazure et al. 1998). Previous large field studies of the photometric
properties of the galaxies in the Coma cluster have been based on
photographic plates (e.g. Godwin, Metcalfe \& Peach 1983; Lugger 1989). CCD studies yield better
photometric precision, but have hitherto been limited to relatively small
areas, mainly focused on the central
regions (e.g. Thompson \& Gregory 1993; Biviano et al. 1995; Bernstein et al. 1995; L\'{o}pez-Cruz et
al. 1997; Secker et al. 1997; Lobo et
al. 1997; de Propris et al. 1998;
Trentham 1998a, 1998b; Mobasher \& Trentham 1998; Andreon 1999). We combine the benefits of wide field coverage with
the photometric accuracy attainable with CCDs, by using the Wide Field Camera (WFC) on
the Isaac Newton Telescope (INT) to image a large part of the Coma cluster
from the dense core to the outskirts of the cluster. Our data offers
the interesting possibility to study the LF for a large range of environments within the cluster. The primary goals of our survey are to study the LF, the colour-magnitude relation and to
complement a large HI survey conducted with the Westerbork Synthesis  
Radio Telescope (WSRT). 

In this paper we present our data and use our catalogue to construct \emph{U},
\emph{B} and \emph{r} band LFs for various regions of the Coma
cluster area. Our study complements previous studies of the LF of the Coma cluster by providing accurate, deep CCD photometry in
three bands, covering an area of 5.2 deg$^2$. Our wide field $U$ band
data is of particular importance, because the $U-R$ colour is a
sensitive indicator of the presence of young stellar
populations. Furthermore, the rest frame $U$ band remains
in the optical window out to $z \sim 1.3$, which makes the $U$ band
observations very useful for comparison to high redshift clusters. Previous $U$ band studies of Coma were limited to small
samples of individual galaxies \cite{bower_etal}. Our
mosaic has similar resolution (in
kpc) and size (in Mpc) as large HST WFPC2 mosaics of high redshift
clusters (e.g. van Dokkum et al. 1998) and provides a valuable low
\emph{z} zeropoint for studies of the evolution of galaxy morphology, colour
and luminosity. For the first time a $U$ band LF for
such a large area of the Coma cluster is determined. In addition we study the dependence of the LF on passband and projected distance from the cluster centre. 

Throughout this paper we use $H_{0}=100~h_{100}$ km s$^{-1}$ Mpc$^{-1}$ and coordinates are given in equinox J2000.0.

\begin{figure}
\center
\includegraphics[width=5cm, angle=270]{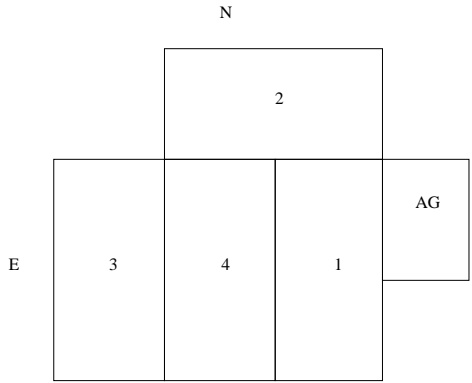}
\caption{Layout of the four science CCDs and the autoguider CCD as used in our survey of the Coma
cluster (rotator angle at 270$\degr$). The four science CCDs cover an area of $34\arcmin \times 34\arcmin$.}
\label{wfc270_layout}
\end{figure}

\begin{figure}
\includegraphics[height=8cm, width=7.8cm]{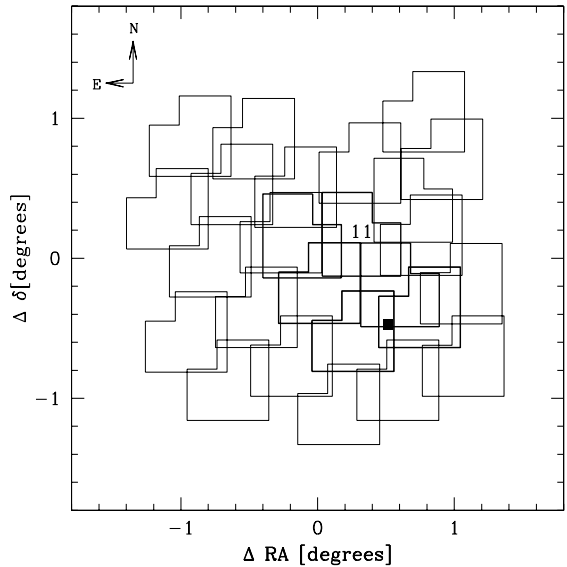}
\caption{Layout of the 25 overlapping pointings covering an area of
$\sim5.2$ deg$^2$. Thick lines delineate pointings which were observed
in the $U$ band, as well as $B$ and $r$. Pointing 11 served as
reference field for the photometric calibration. The position of NGC
4839 is indicated  by the square. Coordinates are given relative
to $\alpha=12^{h}59^{m}43^{s}, \delta=+27\degr58\arcmin14\arcsec$.}
\label{coma_layout}
\end{figure}

\section{Observations}

The data were collected during the nights of March 19--22 1999 with the Wide Field Camera (WFC) on the Isaac Newton Telescope (INT), on Rogue de Los Muchachos on the island of La
Palma (Spain). The WFC consists of four $2048 \times 4100$ pixels EEV CCDs
and a fifth CCD for autoguiding. The layout of the chips is shown in Fig.~\ref{wfc270_layout}. It
is designed to provide a large field survey capability for the prime  
focus of the INT. The sky coverage is $4\times259\arcmin^2$
with a plate scale of $0\farcs333$ pixel$^{-1}$. The camera covers 80
per cent of the unvignetted field of the INT.

 We have imaged a mosaic of 25 overlapping pointings 
covering a total area of $\sim5.2$ deg$^2$ or $2.8 \times 2.8
~h_{100}^{-1}$ Mpc. Broadband filters RGO \emph{U}, Harris \emph{B} and
Sloan \emph{r} were
used for the 6 pointings covering the core and the south-west group
(containing NGC 4839). The other 19 pointings were observed with 
Harris B and Sloan r. For each pointing two exposures, offset by
$\sim 1\arcmin$, were taken. This
ensured that we would be able to determine the relative zeropoint offsets
between exposures and chips and there would be very few gaps in the final mosaic. Furthermore, galaxies in the overlapping regions give us 
good estimates of the errors due to photon noise and flatfielding. Integration times were $2\times300$ s in $B$, $2\times600$ s in $r$ and $2\times900$ s in $U$. The layout of the field is shown in Fig.~\ref{coma_layout}. Thick lines delineate pointings which were observed in the $U$ band, as well as $B$ and $r$.

The overall weather was good with an average seeing during the
first two nights of $\sim
1.4\arcsec$ and $\sim
1.9\arcsec$ during the last two nights in the \emph{r} band. The strategy was to
observe the central parts of the cluster in \emph{U}, \emph{B} and
\emph{r} under photometric conditions. During the first night the standard
Landolt \shortcite{landolt}  fields Sa107-602 and Sa101-427 were also observed at similar
airmasses as the central fields. These standard fields contain many standard stars and give a good spread over the four CCD chips. This
is important for the absolute calibration, since no two chips have  
exactly the same characteristics. In fact it is known that the WFC is
non-linear.

The Coma fields were observed when the cluster had risen to an airmass
$<1.8$. At the start of nights 3 and 4 we took exposures of an empty
field (see Section 4.4). This field was later used to
correct \emph{U} band galaxy counts for foreground/background
contamination.

\section{Initial reduction}

\subsection{Flatfielding}

Prior to any reduction the images were inspected to judge their
quality. For each chip the bias frames per night were checked for
repeating structures and count levels. The bias-structures were also
visible in the science images. We combined all bias frames to
create an average master bias frame for each night and each chip. After several tests
we decided that the bias subtraction had to be performed in two steps;
we
 first subtracted the overscan level by fitting a second order polynomial
to the columns and then subtracted the (overscan subtracted) master bias frames.
 The second step successfully removed the remaining bias structures for chips 1,2 and 4. Chip 3
suffered from a few bad columns. These were removed by interpolation
or by aligning and combining the images taken with a small offset. The rms noise level in the master bias frame was negligible compared to the rms noise in the science frames. 

Accurate flat field images were created from the science frames. 
We have obtained 52 science images of the Coma area per chip per band for the \emph{
B} and \emph{r} bands. For the \emph{U} band we have 23 science images,
including the empty fields. We constructed a \emph{
B} and \emph{r} band masterflat by scaling these images by the mode, rejecting the lowest 10 and highest 15 pixelvalues, followed by median
filtering. The
\emph{U} band masterflats were constructed in the same way, but there we rejected
 the lowest 5 and highest 10 pixelvalues before median filtering.  

As a test we compared our flatfields to the ones of the Wide Field Imaging
Survey which are available from the WFC archive\footnote{http://archive.ast.cam.ac.uk/wfsurvey/wfsurvey.html}. Differences between the flat fields were at the level of $10^{-4}$ or less, assuring that
we have good quality flats.

After flatfielding, approximately 2.5 per cent of the area of chip 3 still
suffers from $\sim 3$ per cent variations in the background, because this
chip is partly vignetted. Because of the overlap between neighbouring
pointings this has very little impact on our final mosaic. On average,
the images are very flat with less than 0.8 per cent variations in the background. 

\subsection{Photometric calibration}
 
The ultimate aim was to obtain
a homogeneous photometric scale over the whole area covered. We dedicated the photometric
anchor point to be chip 4
of pointing 11 (Fig.~\ref{coma_layout}) at airmass 1.000. This field was observed during the
first night under photometric
conditions at a time close to when the standards were taken. First, we determined the photometric offsets between the four chips of the WFC, using objects in overlapping regions between the two offset exposures for each pointing. For the \emph{U} band the number of bright objects in
overlapping areas was low. Therefore, we could not use objects to
determine the offsets, since this would have given unreliable results. Instead, we used the skylevel to determine photometric offsets. First, we made sky-images by filtering out all objects
of the science images in the same way as for the flatfields. Comparison of the mean skylevels of the sky images gave the offsets. The resulting offsets are listed in Table~\ref{offsets}. Errors indicate the uncertainty in the mean of all independent measurements. The $1\sigma$ spread between independent measurements of pairs of different pointings is much larger, and is caused by varying conditions between observations of neighbouring pointings. The offsets depend on the passband, because the response curves of the four CCDs are not identical.

\begin{figure*}
\center
\includegraphics[width=8cm, height=5.4cm]{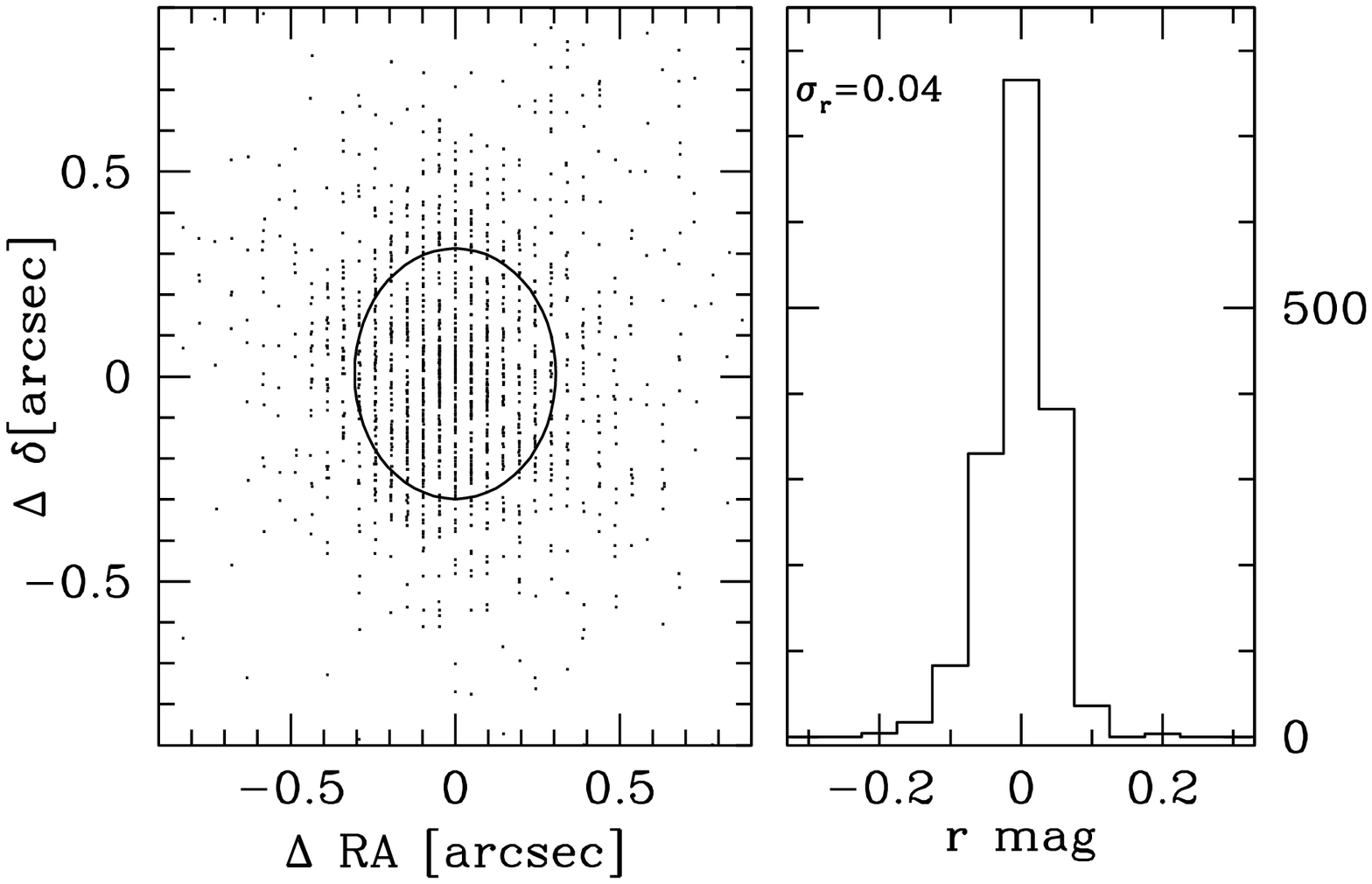}
\includegraphics[width=8cm, height=5.4cm]{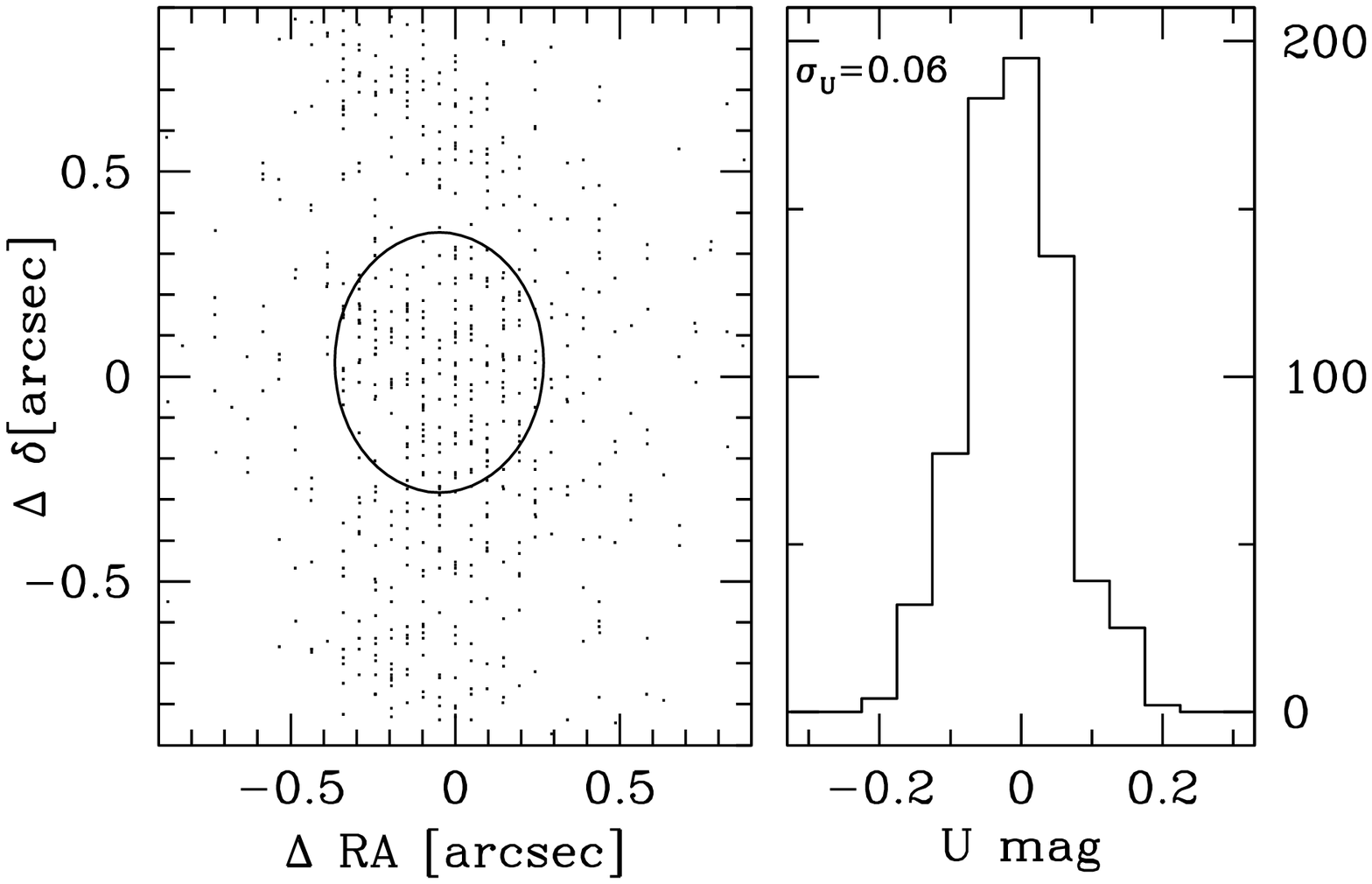}
\caption{Astrometric and magnitude scatter in the $r$ and $U$ bands
for objects in overlapping regions. The circle denotes the astrometric
accuracy $\sigma_{\rm r} \sim0.3 \arcsec$.}
\label{scatterR}
\end{figure*}

We applied these offsets to bring all chips to the same scale. Only the first two nights were photometric and we proceeded by determining the photometric offsets between neighbouring pointings. The
mean offset values are based on the 8 (or more) brightest
objects in the overlapping areas in order to guarantee a reliable determination. We found differences  of a few hundredths of a
magnitude between fields taken at significantly different times (airmasses,
 conditions) in \emph{B} and \emph{r} up to more than a tenth of a magnitude
in \emph{U}. We mapped all the photometric offsets per pointing per
band and scaled all pointings in such a way that the errors were spread
over the total observed region and did not accumulate towards the edges of the covered
area. The final result is that all \emph{B} and \emph{r} band images
have the same zero-point to $\sim{0.04}$ mag and the \emph{U} band
images to $\sim{0.06}$ mag. This is shown for the $r$ and $U$ bands in the right panels of Fig.~\ref{scatterR}.

The photometric
calibration was done using Landolt \shortcite{landolt} standard stars. We did not  
have sufficient standard stars on all the chips to solve for the extinction coefficients
 so we set the extinction coefficients to constant values. We used
average extinction coefficients at the effective wavelengths of the filter+CCD
system, as listed in
Table~\ref{extinction}, to derive the colour terms and zero-points. The zero-points are listed in
Table~\ref{calibration}. The colour terms were found to be very small ($\sim0.01$) and 
are neglected. Since
Coma lies close to the galactic pole the extinction is
insignificant for our purposes (0.043, 0.034 and 0.021 mag in the \emph{U},
\emph{B} and \emph{r} bands respectively). Hence, we did not apply
extinction corrections. 

\begin{table}
\caption{\bf Photometric offsets relative to chip 4}
\begin{tabular}{llll}
chip&U&B$\pm\sigma$&r$\pm\sigma$\\
\hline
1&+0.347$\pm0.015$&+0.337$\pm0.013$&+0.446$\pm0.011$\\
2&+0.170$\pm0.015$&+0.410$\pm0.013$&+0.501$\pm0.017$\\
3&+0.016$\pm0.015$&+0.325$\pm0.014$&+0.376$\pm0.011$\\
\end{tabular}
\label{offsets}
\end{table}

\begin{table}
\caption{\bf Average extinction coefficients for La Palma}
\begin{tabular}{lll}
band&$\lambda_{\rm eff}$ (\AA) &extinction/airmass (mag/airmass)\\
\hline
U&3610&0.45\\
B&4361&0.20\\
r&6216&0.08\\
\end{tabular}
\label{extinction}
\end{table}

\begin{table}
\caption{\bf Zero-points (ZP) of chip 4 for night 1}
\begin{tabular}{ll}
&ZP$\pm\sigma$\\
\hline
U&23.157$\pm0.026$\\
B&24.918$\pm0.031$\\
r&24.822$\pm0.073$\\
\end{tabular}
\label{calibration}
\end{table}

\section{Object catalogues}

The data volume is considerable and requires a fully automated
pipeline for data handling. After the standard reduction steps the processed images were presented
to software
implemented at the Leiden Data Analysis Center (LDAC) for pipeline
processing of overlapping images \cite{deul}. The processing software is a series
of programmes that, when run in a chain, derive source parameters for
any given set of input frames of a given passband. Only the first step is an actual data
\emph{reduction} task in that it reduces the information to be
processed from image data to catalogue data (an object list). All
but the last of the following pipeline routines add information to the
catalogue(s) increasing its size and information content. The final step
merges the catalogues to create a multi-colour object catalogue with
astronomically meaningful source information. The pipeline processing is performed on the full
data set. Along the way we must decide on values for various
parameters. We describe the pipeline steps in more detail below.

\subsection{Detection of objects}

We used SExtractor to
automatically detect objects on each chip individually. The complete
analysis of an image is done in six steps. First, a model of the
sky background is built and parameters describing the global statistics are
estimated. Then the image is background-subtracted, filtered and
thresholded. Detections are then deblended, cleaned, photometered,
classified and written to the final catalogue. For specific details of  
each of these steps the reader is referred to Bertin \&
Arnouts \shortcite{bertin_arnouts} or the SExtractor user's guide.

Objects were identified as the peaks in the (background subtracted)
convolved images that were higher than a given threshold above the
local background. We used most of the standard SExtractor settings except for the
memory parameters which had to be increased in order to detect the
large cD galaxies successfully. The seeing parameter had to be adjusted
for each image individually for a good star/galaxy separation. For all objects positions, magnitudes, basic shape
parameters and star/galaxy (S/G) classifiers were determined and written to a
catalogue. SExtractor produced a catalogue for each image which was
subsequently converted to a format suitable for pipeline
reduction. As a first pipeline processing step, all information of the individual image source extractions
plus all original FITS image header information is combined into one single
output catalogue per band. 

\begin{figure*}
\center
\includegraphics[height=7cm, width=15cm]{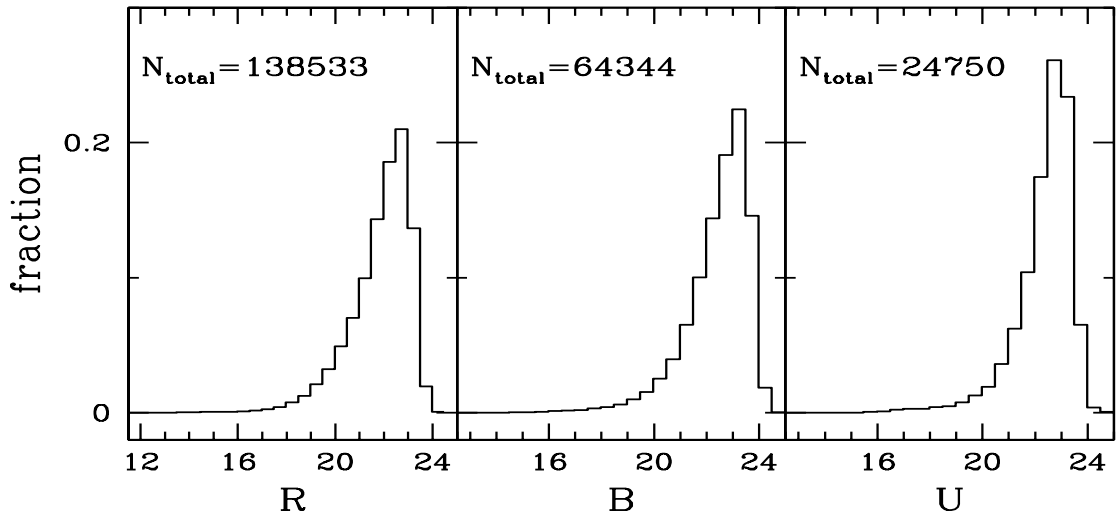}
\caption{Histograms of the normalized raw \emph{r}, \emph{B} and \emph{U} band number
counts. N$_{\rm total}$ gives the
total number of catalogue entries for each passband.}
\label{lfhistos}
\end{figure*}

\subsection{Astrometry}

Astrometric calibration was performed by pairing the input position
catalogue (USNO-A2) with the extracted object information. For
multiple band processing intercolour pairing is done first and between
frames in overlap, overlap pairing is performed as well. The derived
astrometric solution is then applied to all the objects in the set of
frames. Sky coordinates (RA and Dec), as well as corrected geometric
parameters are calculated. The final precision of this calibration
depends on the accuracy of the source extractions, input catalogue accuracies
and the correctness of the functional description of the
distortions. We find $\sigma_{\rm r}\sim0.3\arcsec$ as shown in the left panels of Fig.~\ref{scatterR}.

\subsection{Final object catalogue}

The final step in creating a catalogue containing useful multi-colour
astronomical data is to merge all catalogues to get one source per
position on the sky. All information referring to the same astronomical object is
gathered and merged. Position information is a weighted mean of all
detections where the weighting is based on detection signal-to-noise
and detection environment conditions. Details of the merging are
configurable. When the brightness contrast between overlapping
galaxies is too low the software is not able to deblend them
correctly, resulting in an erroneous catalogue entry. On the other
hand, setting the deblend contrast parameter too low causes SExtractor
to consider bright star forming regions in a galaxy as separate
objects. Then, there is the possibility that deblended objects are
merged by the pipeline software when creating the final object catalogue. This can happen in cases where objects are small compared
to the errors in position and shape parameters. We carefully inspected
dense regions in our mosaic and conclude that erroneous catalogue
entries are rare and consequently do not affect our results.

\begin{figure*}
\center
\includegraphics[height=16cm, width=16cm]{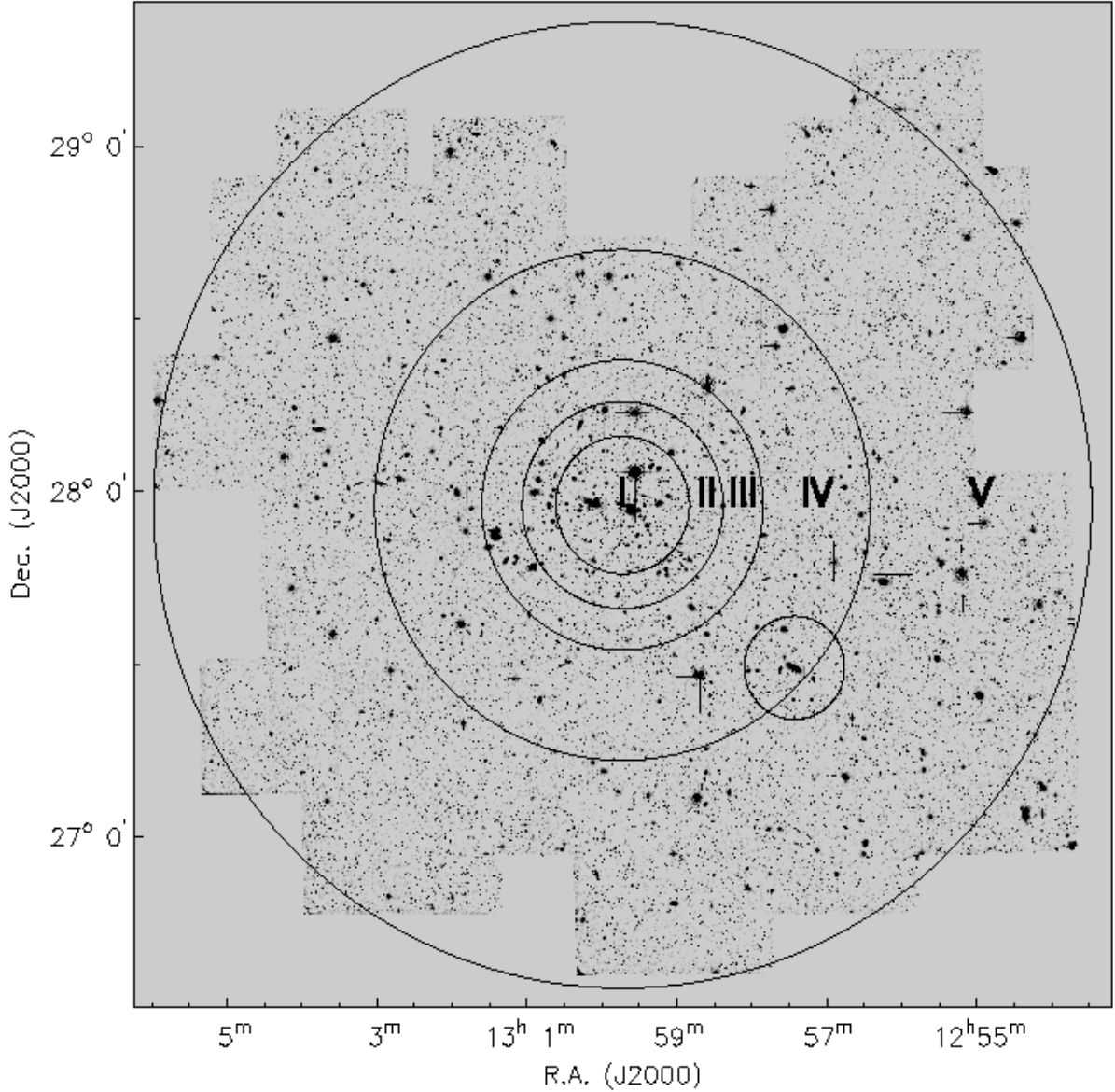}
\caption{Overview of the total area covered in the \emph{B} and \emph{r}
band with annuli overlaid. Radii are: I: 0-0.2, II: 0.2-0.3, III:
0.3-0.42, IV: 0.42-0.74 and V: 0.74-1.4 degrees ($1\degr=1.22
~h_{100}^{-1}$ Mpc at Coma distance). The circular area centered on NGC 4839 has a radius of $0.15\degr$.}
\label{circle_layout}
\end{figure*}

Star/galaxy separation was performed based on SExtractor's
stellarity index. For bright objects stars and galaxies are easily separated, but towards fainter magnitudes and for bad seeing the division is not so clear. For most
purposes we consider the objects
with a S/G classifier value smaller than 0.8 in the \emph{r} band
(best seeing) to be galaxies. Bright (saturated) stars tend to have a
S/G classifier smaller than this. We filtered these out by demanding a
S/G classifier $<0.1$ for the brightest magnitudes. We visually
inspected all bright objects to conclude that there is no
contamination by stars up to at least $m_{\rm r}=15$. 
Beyond $m\sim19$ mag
we probably still have a fraction of stars in the sample, but for most purposes it is better to be contaminated
by a (small) fraction of stars than to reject compact
galaxies. Histograms of the catalogue's raw number counts are shown in
the panels of Fig.~\ref{lfhistos}. Troughout this paper we use
SExtractor's MAG\_BEST as
magnitude estimator. The final mosaic, composed of
all pointings observed in the $B$ and $r$ bands, is shown in Fig.~\ref{circle_layout}.

\subsection{Control fields}

For a survey as large and deep as ours, foreground/background
subtraction can only be treated statistically, since only the
brightest galaxies have measured redshifts. The main source of
uncertainty in the determination of LFs comes from number statistics
and background variance. Usually, flanking fields are used to get an estimate of the
foreground/background correction and its variance. We expect
that over the large area of the WFC the effects of cosmic variance are not important.

For the \emph{U} band we have used $4\times900$ s observations of an empty field located at $\alpha=8^{h}00^{m}00^{s},
~\delta=+50\degr00\arcmin00\arcsec$ spanning $\sim980$
arcmin$^{2}$. For both the \emph{B} and \emph{r} bands we have made use of
the Wide Field Survey (WFS) data archive to get images of random fields. The \emph{B} band control images are eight 600 s
exposures, spanning $\sim2000$ arcmin$^{2}$ and centered at
$\alpha=12^{h}55^{m}55^{s}, ~\delta=+27\degr01\arcmin40\arcsec$ and
$\alpha=12^{h}53^{m}40^{s}, ~\delta=+26\degr20\arcmin00\arcsec$. The
\emph{r} band control images are four 600 s exposures, spanning $\sim1000$
arcmin$^{2}$ centered at $\alpha=16^{h}04^{m}26^{s}, 
~\delta=+54\degr49\arcmin59\arcsec$. For the \emph{B} and \emph{r} band control fields we rely on the (photometric) reduction
of the four chips by the WFS team. 

All control images were pipeline reduced to give control catalogues which were
filtered with the same criteria as the Coma fields. We applied relative extinction corrections \cite{schlegel} to bring the extinction in the control fields
into agreement with the Coma extinction, not zero extinction. 

\begin{table}
\caption{\bf Total cluster counts and field counts estimates per deg$^2$}
\begin{tabular}{cllllll}
mag&$r_{\rm total}$&$B_{\rm total}$&$U_{\rm total}$&$r_{\rm field}$&$B_{\rm field}$&$U_{\rm field}$\\
\hline
11.23          & 0&           0  &   0      & 0  &0 & 0\\
       12.23  &1.54 &0 &0      &   0 &          0  &         0\\
       13.23      &6.54 & 0.38 &0   &  0  &      0     &      0\\
       14.23      &19.42 &0.76&0.73  &   0   &     0    &      0\\
       15.23     &33.85 &11.92  &5.10 &  10.44    &    6.96    &       0\\
       16.23    &59.04 &33.08  &45.14 &  27.84    &   10.44     &  0\\
       17.23    &142.1 &52.5 &100.5 &  59.16    &   15.66    &   0\\
       18.23    &415.2  &108.3  &144.9 &  302.7    &    60.9    &   25.7 \\
       19.23    &1127 &245     &274.5 &   1044    &   200.1    &   198.2\\
       20.23    & 2634  &631.3 &752.1 &   2551   &    492.4    &   561.7\\
       21.23     &5444  &1628   &2294&   5282  &      1529    &    1733\\
\end{tabular}
\label{fieldcounts}
\end{table}

\begin{figure}
\center
\includegraphics[height=8cm, width=8cm]{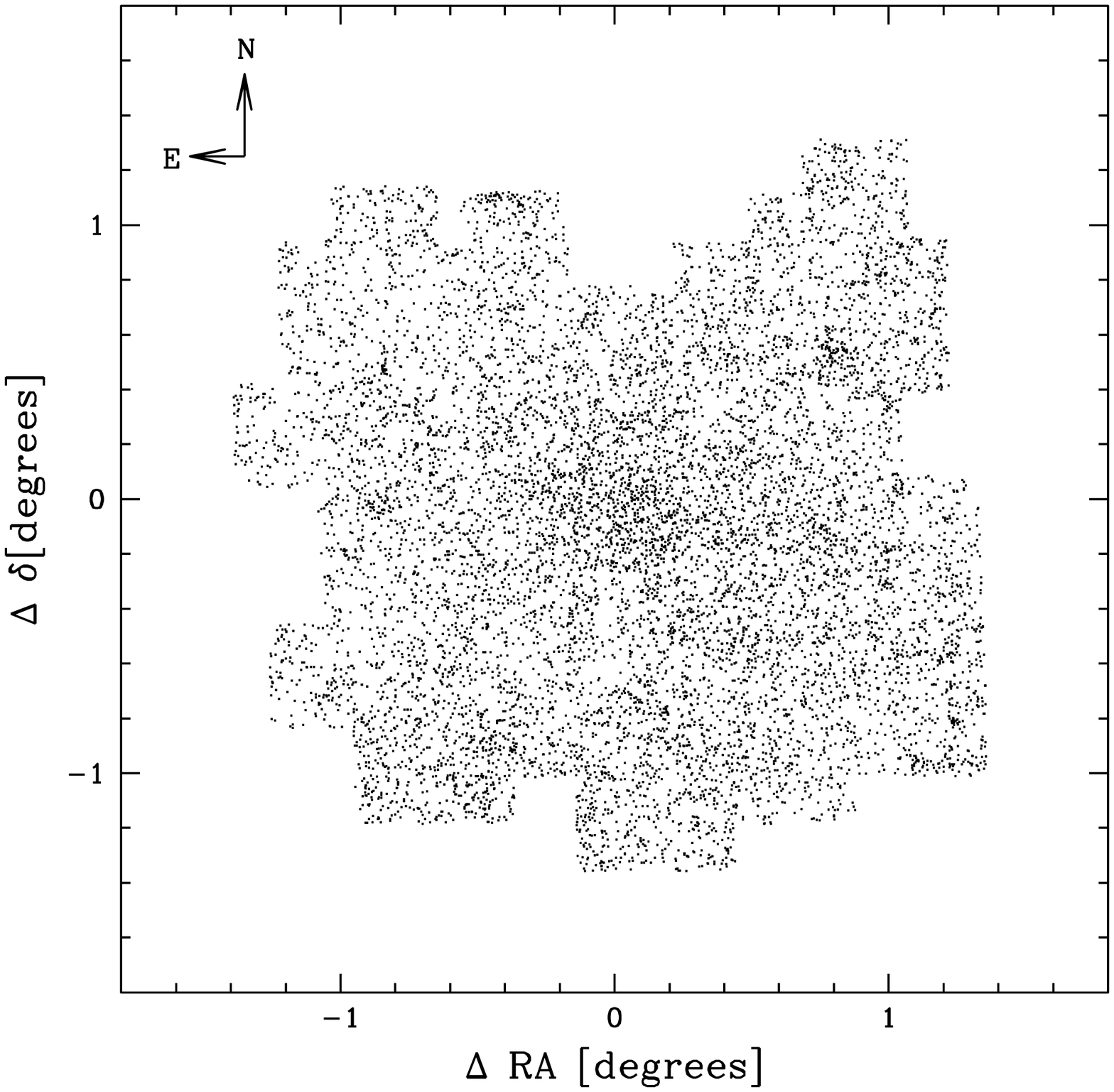}
\caption{Distribution of objects with $m_{\rm B} < 21.73$. Coordinates are given relative to
$\alpha=12^{h}59^{m}43^{s},
\delta=+27\degr58\arcmin14\arcsec$.}
\label{skycov}
\end{figure}

\subsection{Completeness}

The histograms shown in Fig.~\ref{lfhistos} suggest that the limiting
magnitude is $\sim 22.5$ mag for all bands. However, due to less than
perfect conditions in the last two nights this limit does not apply to
the total observed area. To have uniform completeness for the total mosaic we compute LFs up to $m_{lim} = 21.73$. The positions of these objects, relative to the cluster centre, are plotted in Fig.~\ref{skycov}. The cluster is visible as a density enhancement on top of a uniform background. We have only considered clean detections (with good extraction
flags) with more than 5 connected pixels above the background
threshold. 

Detection limits in
general depend on the interplay between scale length/effective radius,
magnitude and
inclination, i.e. surface brightness. Edge-on galaxies of a certain
magnitude are easier to detect than their face-on counterparts. Deep surveys have shown the
existence of galaxies with surface brightnesses fainter than the
night sky. These low surface brightness (LSB) galaxies are more
likely to be missed at a given magnitude and inclination than their high surface
brightness counterparts. Most of the LSB galaxies investigated in any detail
are either late-type and disk dominated (de Blok et
al. 1995; McGaugh \& Bothun 1994), or giant,
Malin-1-like galaxies (Sprayberry et al. 1995; Pickering
et al. 1997). 

\begin{figure}
\center
\includegraphics[width=8cm, height=8cm]{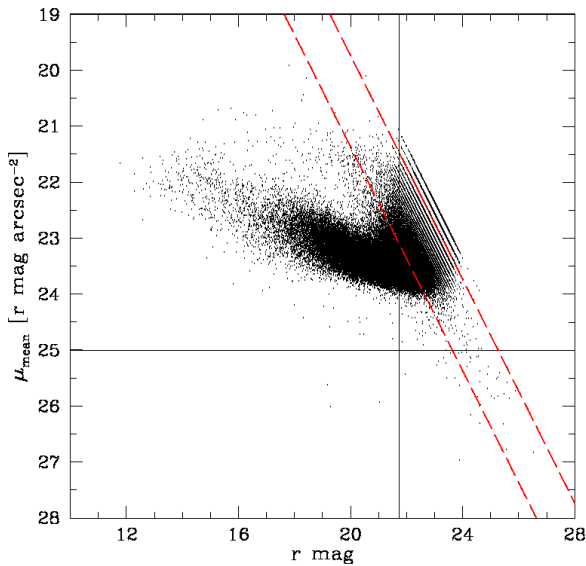}
\caption{Luminosity versus mean surface brightness plot. The
selection boundaries are indicated by the solid lines. The dashed
lines represent the minimum size detection lines for a seeing of
$1.5\arcsec$ and $1\arcsec$ (from left to
right).}
\label{muM}
\end{figure}

In order to estimate these effects we followed two approaches. First
we generated a $r$ band luminosity-surface brightness plot for all
objects (mostly galaxies) in
our catalogue. The data in our catalogue do not contain estimates of
the central surface brightnesses, but we can compute mean surface
brightnesses for all objects based on total flux and area above the
analysis threshold. In
Fig.~\ref{muM} we show the resulting luminosity-surface brightness
plot along with the relevant magnitude and surface brightness
selection boundaries we used. The
dashed lines represent the luminosity-surface brightness relations for
objects with the minimum detectable size (the seeing disk). We show
minimum size detection lines for a seeing of $1.5\arcsec$ and
$1\arcsec$, from left to right, and it can be seen that occasionally
the seeing drops even below $1\arcsec$. The mean isophotal limits for the $U$ and
$B$ bands are 25.10 and 25.42 mag arcsec$^{-2}$ respectively. From this figure it is evident that we are
not likely to miss galaxies due to their low surface
brightness. Second we generated artificial
elliptical galaxies with de Vaucouleurs' law \cite{de_vaucouleurs} light profiles and spiral
galaxies with exponential
light profiles. We then added these galaxies into empty regions of our Coma
images and verified whether SExtractor could recover these using the
same selection criteria as for Coma fields. From our simulations we
estimate that we are able to detect ellipticals with effective radii $r_{\rm
e}\sim3 ~h_{100}^{-1}$ kpc or $r_{\rm e}\sim9\arcsec$ down to $\sim19.5$,
$\sim19.5$ and $\sim19$ mag in the
\emph{U}, \emph{B} and \emph{r} band respectively. Furthermore, we are able to detect 
dwarf galaxies modelled as exponential disks with $h_{\rm
d}\sim1$
kpc down to $\sim19.5$, $\sim20$ and
$\sim20$ mag or $\mu_{0}=23.8,~24.3$ and 24.3 mag
arcsec$^{-2}$ in the \emph{U},
\emph{B} and \emph{r} band respectively. At the distance of Coma,
$m=20$ corresponds to $M=-14.2$. Disk dominated LSB galaxies typically have
$h_{\rm d}\sim3$ kpc and $\mu_{0,\rm B}\sim23.2$ mag arcsec$^{-2}$. Assuming typical
colours as in the de Blok et
al. \shortcite{de_blok} sample, this corresponds to $\mu_{0,\rm
U}\sim23.1$ and $\mu_{0,\rm
r}\sim22.4$ mag arcsec$^{-2}$. Our simulations show that such LSB galaxies are within our detection
limits. 

\begin{figure*}
\includegraphics[height=23cm, width=16cm]{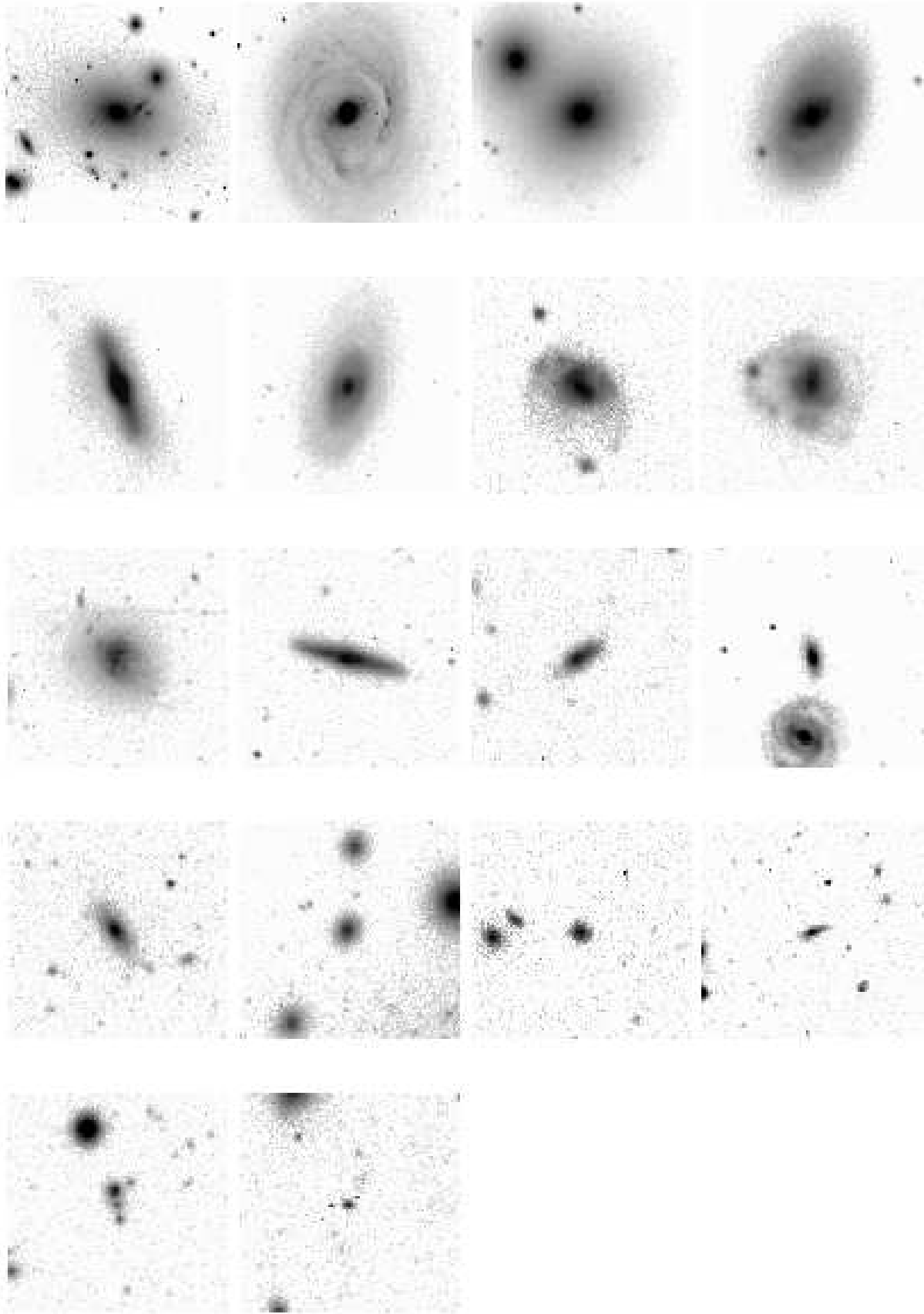}
\caption{Examples of galaxies drawn from our catalogue for
magnitude bins $M_{\rm B}$= -21 to -13 in steps of one magnitude, two
examples per bin.}
\label{examples}
\end{figure*}

To illustrate the data quality we show examples of galaxies in
Fig.~\ref{examples}. These have been drawn from our catalogue for
magnitude bins $M_{\rm B}$= -21 to -13 in steps of one magnitude, two
examples per bin.

\begin{table}
\caption{\bf Best-fitting Schechter parameters for the luminosity functions of
the complete sample}
\setlength{\tabcolsep}{1.2mm}
\begin{tabular}{lcccr}
Filter&$M^*$&$\alpha$&$\phi^*$[$h^2_{100}$ Mpc$^{-2}$ mag$^{-1}$]&$\chi^2_{\nu}$\\
\hline
&&&\\
U&$-19.39^{+0.33}_{-0.40}$&$-1.54^{+0.036}_{-0.030}$&$11.7\pm1.05$&27.8\\
&&&\\
B&$-19.09^{+0.36}_{-0.40}$&$-1.32^{+0.056}_{-0.049}$&$13.7\pm0.02$&5.35\\
&&&\\
r&$-20.63^{+0.26}_{-0.34}$&$-1.22^{+0.034}_{-0.036}$&$20.44\pm0.48$&2.87\\
&&&\\
\end{tabular}
\label{totalparameters}
\end{table}

\begin{figure*}
\center
\includegraphics[height=8cm, width=16cm]{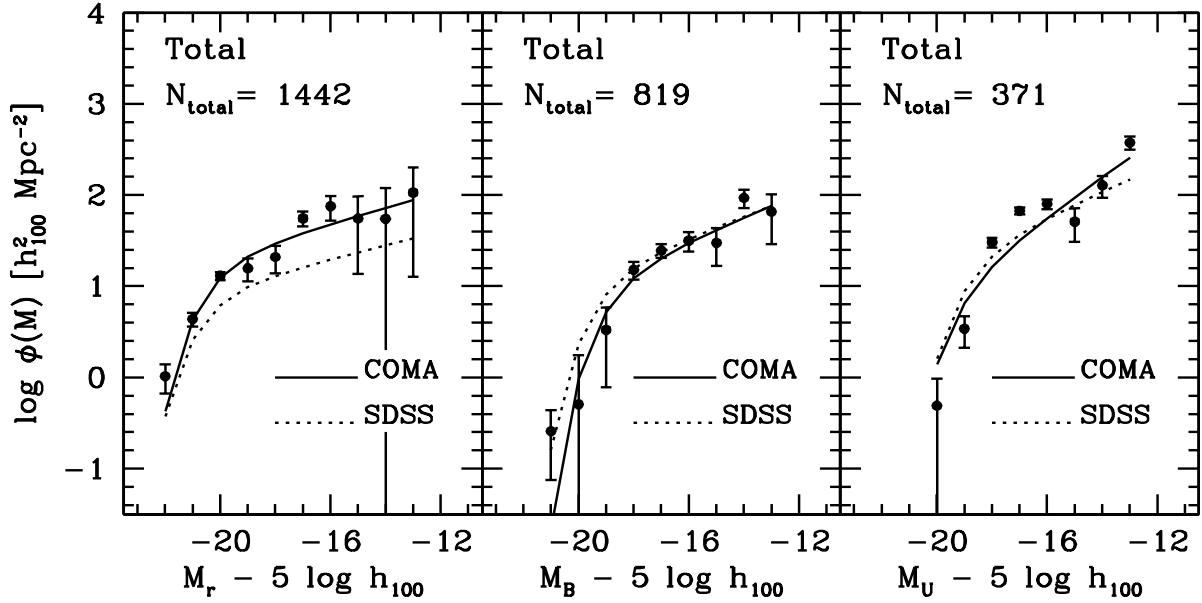}
\caption{LFs for the total area observed. The solid lines represent
best-fitting Schechter functions with parameters as given in
Table~\ref{totalparameters}. The dashed lines correspond to the Sloan
Digital Sky Survey field LFs. For clarity the SDSS LFs were
renormalized by adding 2.8 to log $\phi(\rm M)$ in all bands. N$_{\rm total}$ gives the estimated
number of Coma galaxies up to -15.2.}
\label{totallfs}
\end{figure*}

\begin{figure*}
\center
\includegraphics[height=8cm, width=16cm]{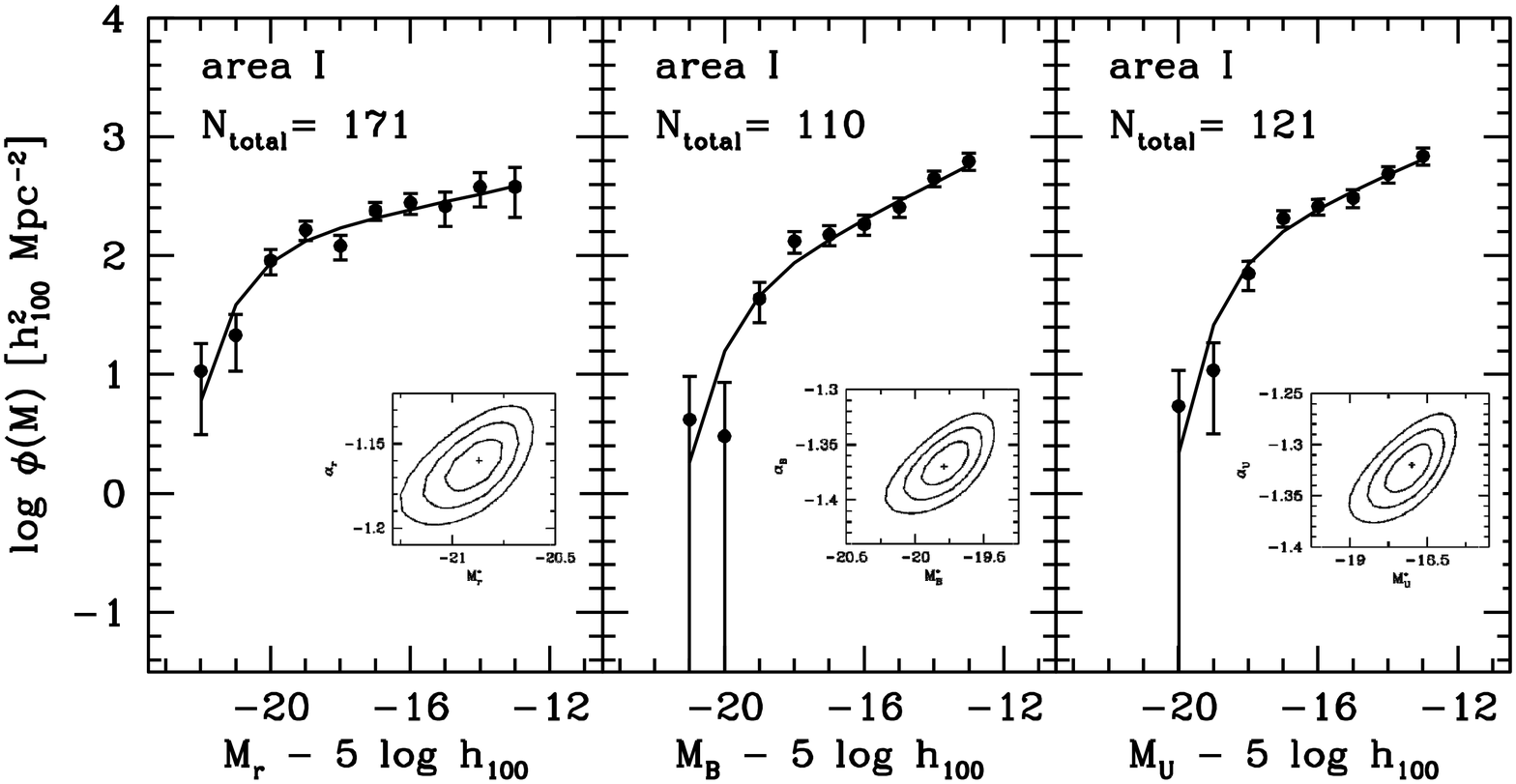}
\caption{LFs for area I. Solid lines represent the best-fitting Schechter
functions with parameters as listed in Table~\ref{centralparameters}. The insets show the 1, 2 and 3$\sigma$ contour levels of the
best fitting Schechter function parameters. N$_{\rm total}$ gives the estimated
number of Coma galaxies up to -15.2.}
\label{corelfs}
\end{figure*}

\section{Total luminosity functions}

We constructed foreground/background corrected LFs by subtracting counts in the control catalogues from the Coma counts. From Fig.~\ref{lfhistos} it is obvious that the number of
foreground/background counts is a strongly varying function of
magnitude. Especially the faintest points of the LFs are largely
influenced by an inaccurate determination of the amount of
contaminating galaxies. We used very large control fields, and large
bin-widths compared to the photometric uncertainties. Therefore,
effects caused by errors in the determination of foreground/background
counts will be suppressed. Furthermore, we carefully calculated
effective areas for all control fields and the Coma mosaic by counting
unblotted pixels on each frame. Pixels which reduced the area for
object detection, e.g. dead columns, saturated stars etc., were
blotted prior to the pixelcounting. The total cluster counts and the field counts estimates are listed in Table~\ref{fieldcounts}. 

In Fig.~\ref{totallfs} the LFs for the complete data set are shown. Solid lines represent best-fitting Schechter
functions with parameters as listed in
Table~\ref{totalparameters}. The faint end slopes increase towards
shorter wavelengths. The best-fitting Schechter functions are
rather poor representations of the data for all bands, with several
points lying more than $1\sigma$ away from the best-fitting values. The
$\chi^2$ statistic confirms the eye's impression that the LFs for the
complete data set cannot be represented by single Schechter functions. These are the first accurate determinations of LFs for such large areas of a cluster. Furthermore, to our knowledge the \emph{U} band LFs presented here are the first ever published for the Coma cluster. 

\begin{table}
\caption{\bf Schechter parameters for the central luminosity functions}
\setlength{\tabcolsep}{1.2mm}
\begin{tabular}{lcccrcccr}
Filter&$M^*$&$\alpha$&$\phi^*$[$h^2_{100}$ Mpc$^{-2}$ mag$^{-1}$]&$\chi^2_{\nu}$\\
\hline
&&&\\
U&$-18.60^{+0.13}_{-0.18}$&$-1.32^{+0.018}_{-0.028}$&$135.1\pm4.3$&1.78\\
&&&\\
B&$-19.79^{+0.18}_{-0.17}$&$-1.37^{+0.024}_{-0.016}$&$61.7\pm2.4$&1.71\\
&&&\\
r&$-20.87^{+0.12}_{-0.17}$&$-1.16^{+0.012}_{-0.019}$&$130.3\pm11.6$&1.34\\
&&&\\
\end{tabular}
\label{centralparameters}
\end{table}

\section{Dependence of luminosity functions on radial distance from the cluster centre}
 
In order to study the dependence of the LF on radial distance from the cluster centre we defined five areas as annuli with varying widths and radii projected on the cluster centre, as shown in Fig.~\ref{circle_layout}. For each of these areas we carefully determined the effective area for galaxy detection and constructed the corresponding LFs. The contamination by foreground/background galaxies becomes severe towards the outskirts of the cluster. Their numbers become equal or greater than the Coma counts at $r>19$, $r>18$ and $r>17$ in annulus I, III and IV, respectively. 

\begin{figure*}
\center
\includegraphics[width=8cm, height=8cm]{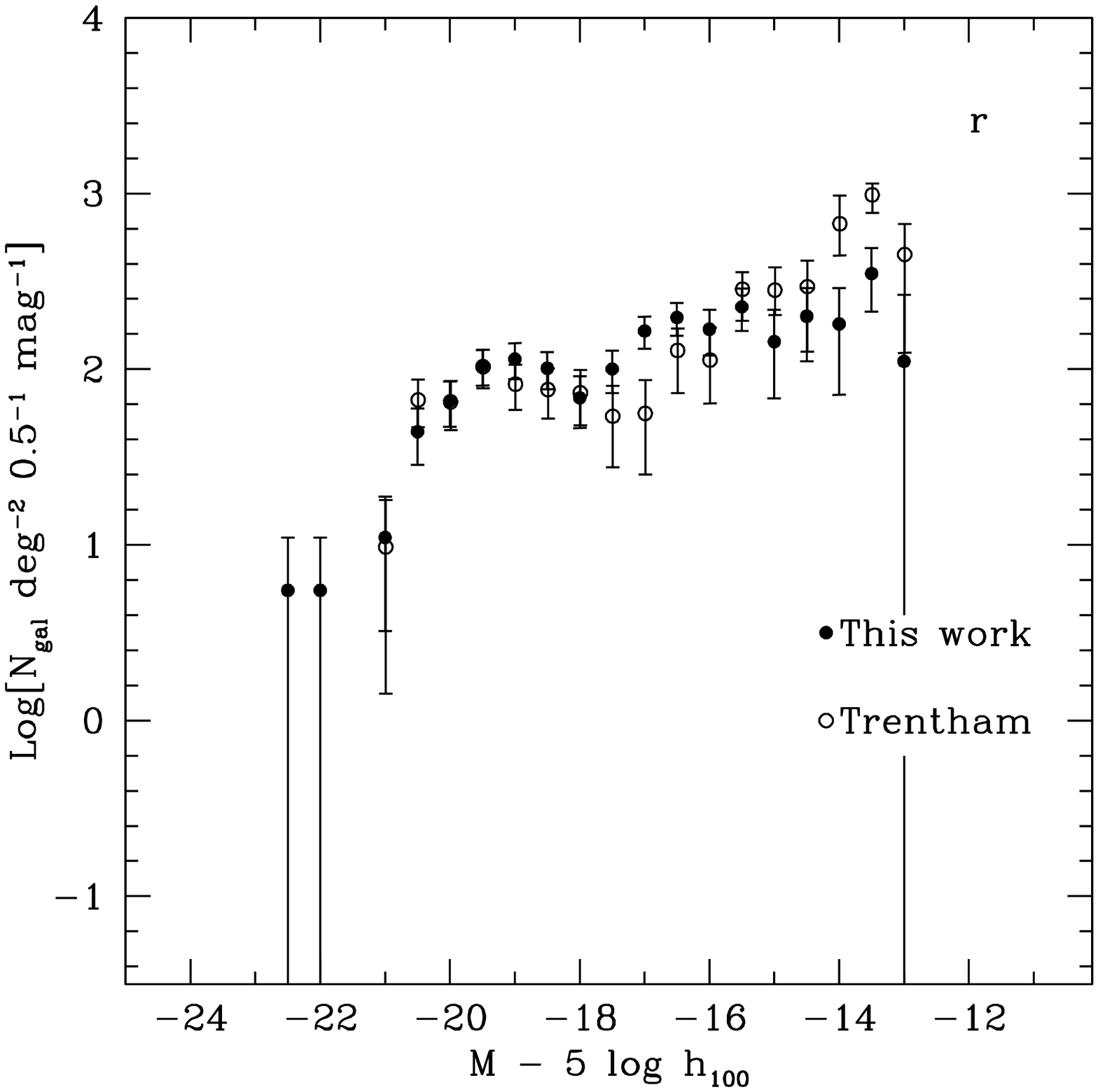}
\includegraphics[width=8cm, height=8cm]{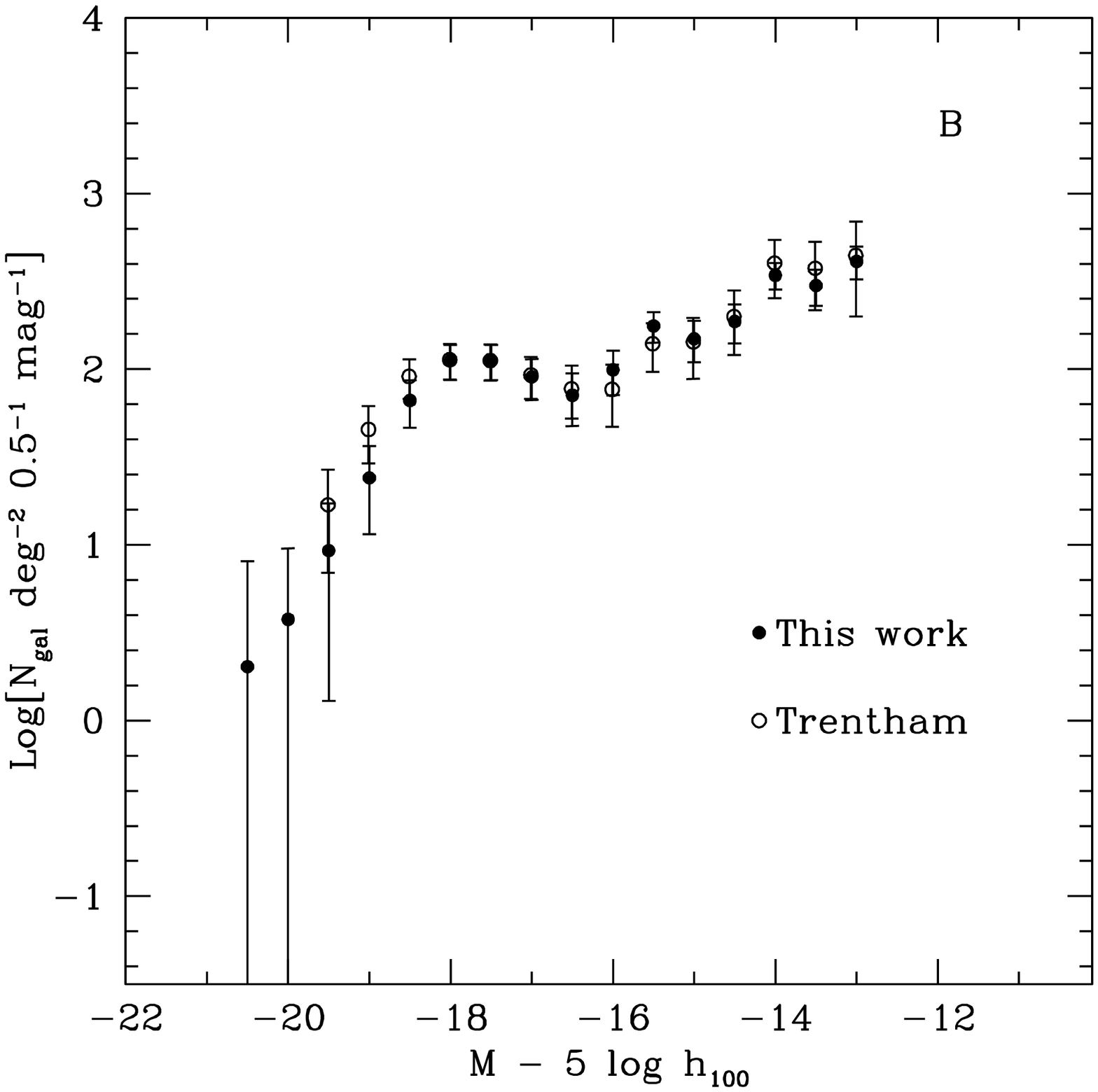}
\caption{Comparison of our data to those of Trentham
\shortcite{trentham1}, restricting our data to the same area covered
by the Trentham \shortcite{trentham1} survey. There is very good
agreement between the two independent determinations of the Coma LF in
this small region.}
\label{trentham}
\end{figure*}

We arbitrarily defined the core of the Coma cluster as the area with $r<245
~h_{100}^{-1}$ kpc (area I). This is comparable in size to the \emph{total} observed areas in previous studies. In Fig.~\ref{corelfs} we show the central LFs with best-fitting Schechter functions
overplotted as solid lines. The insets show the 68, 95 and 99 per cent
confidence levels for the $M^*$ and $\alpha$ parameters resulting from the fit to the binned data. The best-fitting parameters are listed in Table~\ref{centralparameters}. The bright end of the \emph{B} band central LF is not adequately represented by the Schechter function. The faint end, however, is well represented by the Schechter fit. We do not confirm the dip, located at
$M_{\rm B}\sim-17.2$, reported by Biviano et al. \shortcite
{biviano_etal}. We stress that our LF has been determined with
completely different data and methods, complicating any direct comparison of results. Their estimate of the faint end slope ($\alpha_{\rm B}=-1.3\pm0.1$) is, however, in agreement with our result. Comparison of our \emph{B} band LF with the typical richness
class 2 composite cluster LF \cite{trentham2} shows that the faint ends are consistent up to the completeness limit. Beyond the completeness limit the composite LF rises more steeply than the
LF we have derived. Our value of the faint end slope of the central \emph{r} band LF
is in agreement with Lugger \shortcite{lugger} ($\alpha_{\rm R}\sim
-1.19\pm0.17$), but somewhat shallower than the slopes derived for Coma by e.g. Bernstein et al. \shortcite{bernstein_etal},
L\'{o}pez-Cruz et al. \shortcite{lopez-cruz_etal} and Secker et
al. \shortcite{secker_etal} who all find $\alpha_{\rm R}\sim -1.4$. We stress that a direct comparison is hampered by the fact that
these authors use different areas or composite LFs. To demonstrate
more clearly that the LF is mainly shaped by the area we extracted
Trentham's \shortcite{trentham1} area and constructed $B$ and $r$ band LFs. In
Fig.~\ref{trentham} we compare the results of both determinations of the LFs. Up to the completeness limits the LFs are very similar
showing that not the methodology employed, but the area chosen has
the most crucial effect on the
shape of the LF.

\begin{figure*}
\center
\includegraphics[height=16cm, width=16cm]{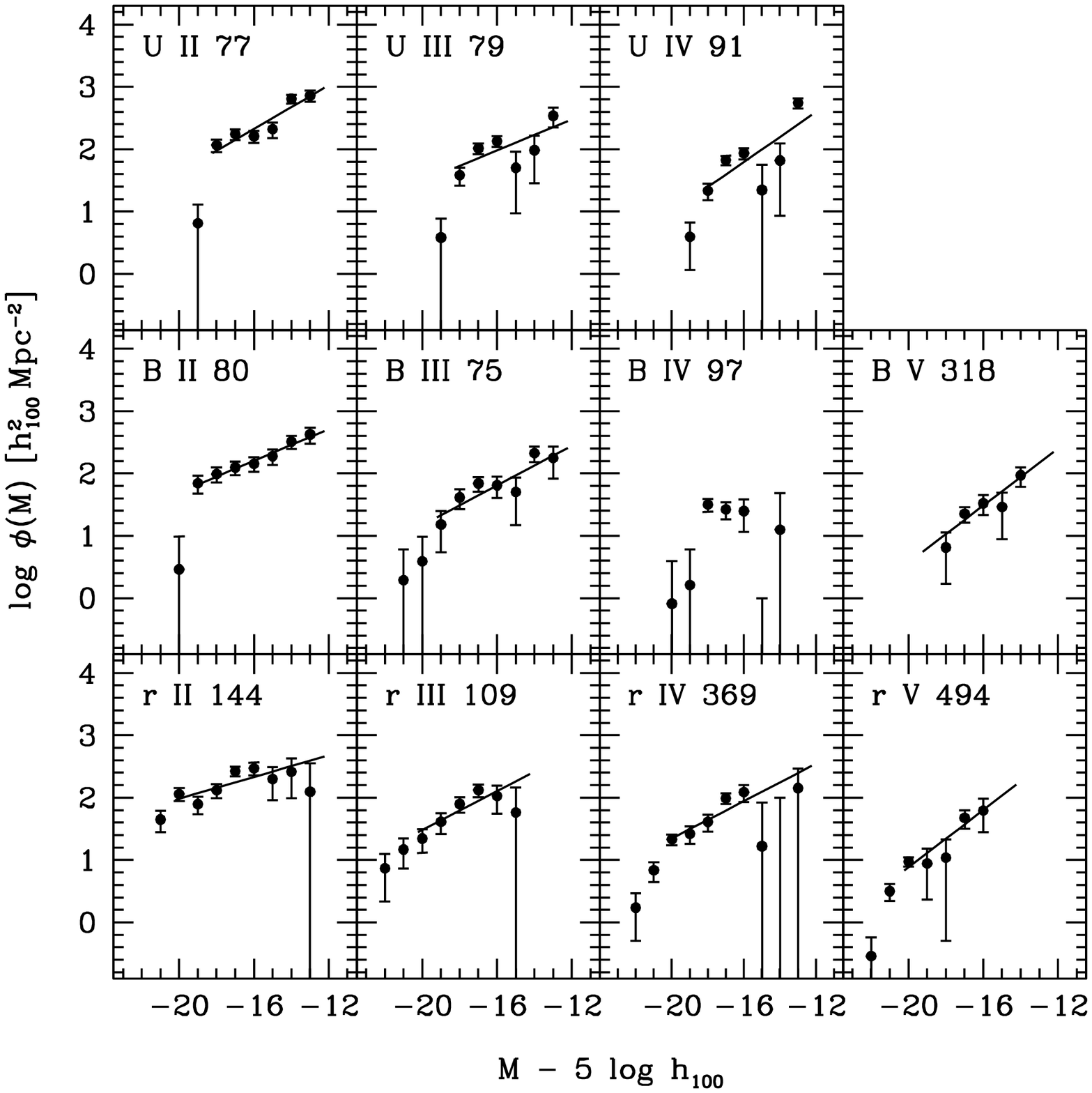}
\caption{LFs for the annuli of Fig.~\ref{circle_layout}. In the top of
each figure we indicate: filter, annulus number and estimated number
of Coma galaxies up to -15.2. The lines represent the fits to the
faint ends.}
\label{lfmosaic}
\end{figure*}

In general a single Schechter function is a reasonable
representation of all the central LFs with 2 points lying $1\sigma$ or
more from the best-fitting value for all bands. It is expected from LF studies of \emph{composite} dense and
loose clusters and of single rich and poor clusters that the LF shape
depends on environment. Below, we study the change in the shape of the LF
as function of position \emph{in} the
cluster. We will examine whether differences can be attributed to
effects of the local environment.

 In the panels of Fig.~\ref{lfmosaic} the LFs corresponding to the
 annuli II to V of Fig.~\ref{circle_layout} are shown for all  
 bands. It is clear that these LFs are not simply scaled versions
of the central LFs.

The $U$ band LFs are sensitive to star forming galaxies, and are therefore
a poor indicator of the underlying mass distribution. They show significant curvature and at some
radii dips, as reported for other bands (e.g. Biviano et
al. 1995; Andreon \& Pello 1999). Because of these dips these LFs cannot be adequately fitted by single Schechter functions. 

The $B$ band LF of annulus II still resembles the central LF, but in
annuli III and IV the faint end behaves differently. Even further
out, in annulus V, the galaxies with $M_{\rm B} \sim-19$ or brighter
become very rare and the faint end seems to steepen again, but only marginally. 

The $r$ band LF of annulus II is relatively flat. At larger radial
distances from the cluster centre the LFs become much steeper. 

\begin{figure}
\includegraphics[height=8cm, width=8cm]{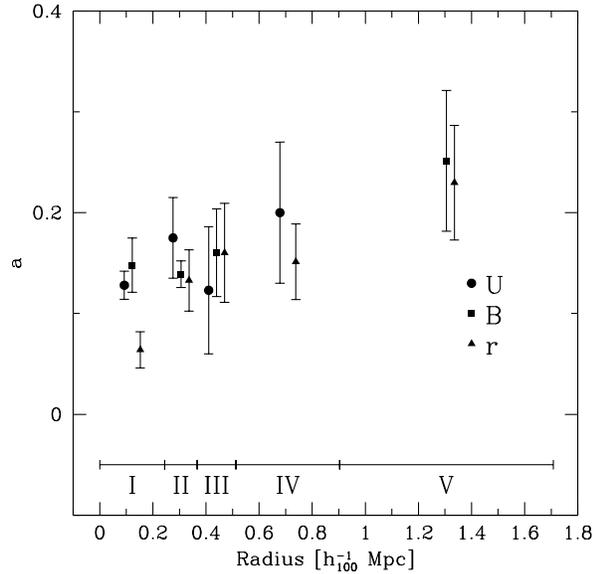}
\caption{Power law slopes $a$ as function
of cluster radius. The faint ends of the LFs become somewhat steeper towards the cluster outskirts.}
\label{alpha}
\end{figure}

We quantified these trends as follows. We fitted a power law function
($b~10^{aM}$) to the faint ends of the LFs in order
to study their dependence on radial distance from the cluster
centre. The LFs were fitted for $M_{\rm U}>$ -18, $M_{\rm B}>$ -19 and $M_{\rm
r}>$ -20, respectively. In Fig.~\ref{alpha} we plot the power law slopes as
function of cluster radius. We omitted the slope of the $B$ band LF for area IV: the
value of the slope is extremely sensitive to the fitting region,
and hence not well constrained. In general, the faint end slopes become steeper
towards larger cluster radii.

\begin{figure}
\includegraphics[height=8cm, width=8cm]{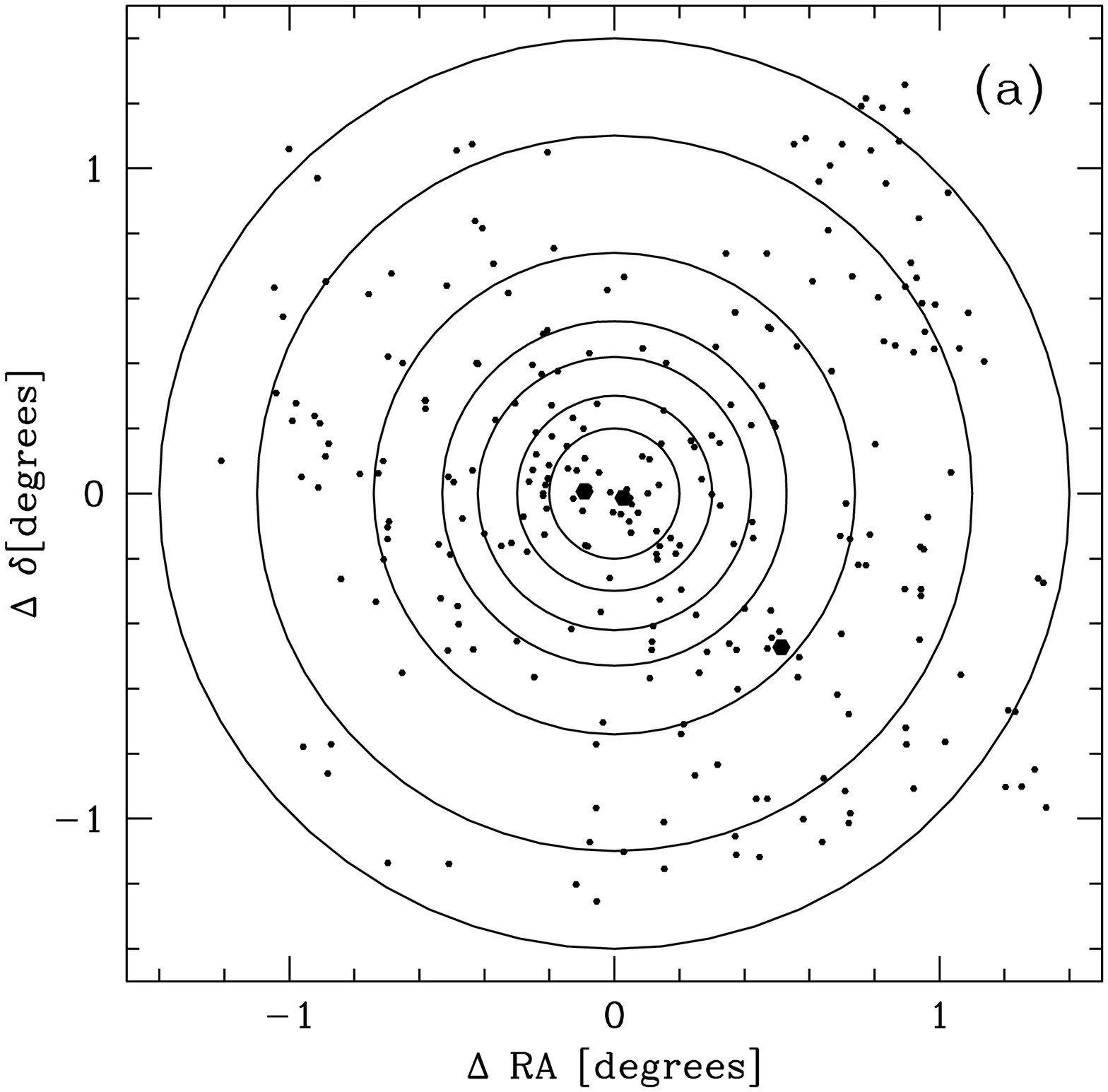}
\includegraphics[height=8cm, width=8cm]{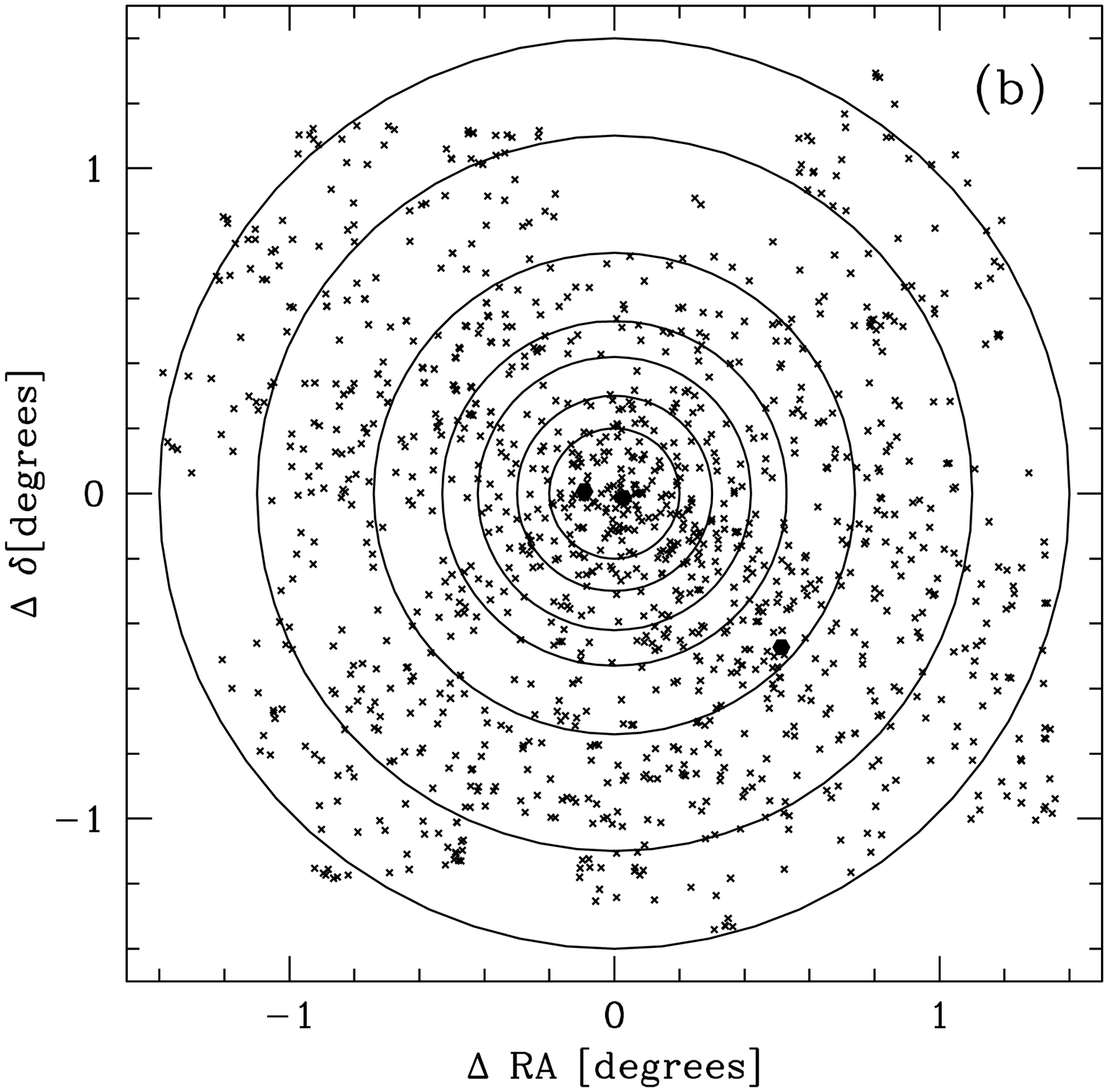}
\caption{(a) Filled hexagons represent galaxies with $-23.5<M_r<-19.5$ (giants). The 3 large hexagons represent NGC 4889, NGC 4874 and
NGC 4839. (b) Crosses
represent galaxies with $-19.5<M_r<-16.5$ (dwarfs). Overplotted are
annuli with radii: 0-0.2, 0.2-0.3, 0.3-0.42, 0.42-0.53, 0.53-0.74,
0.74-1.1 and 1.1-1.4 degrees ($1\degr=1.22
~h_{100}^{-1}$ Mpc at Coma distance).}
\label{coverage}
\end{figure}

\begin{figure}
\includegraphics[height=10cm, width=8cm]{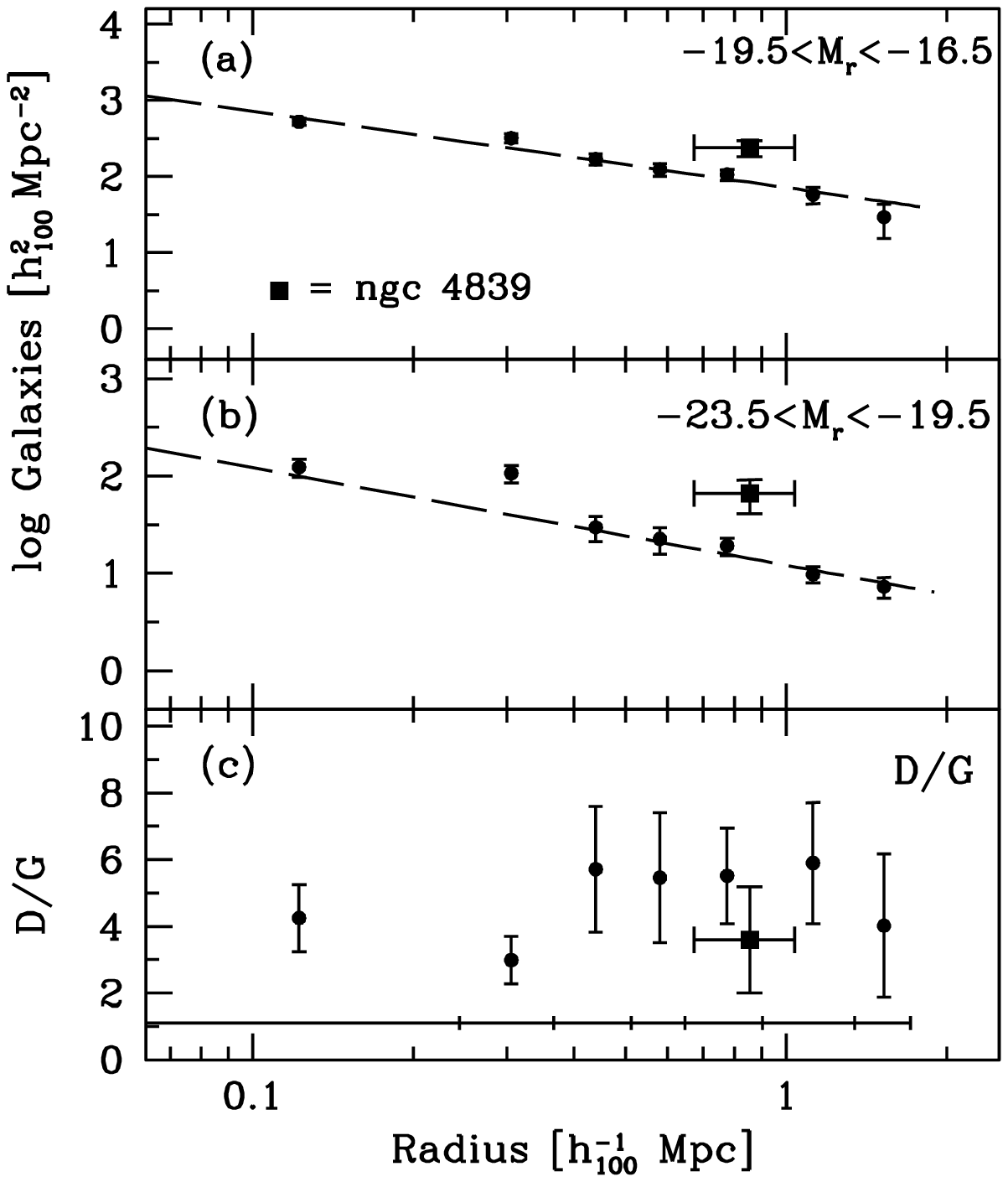}
\caption{(a) The number density of dwarf galaxies as function of
distance from the cluster centre. (b) The number density of giant galaxies as function of
distance from the cluster centre. (c) The dwarf-to-giant
ratio (D/G) as function of distance from the cluster centre. The NGC 4839 group is indicated by a square. Dashed
lines correspond to isothermal profiles. Horizontal errorbars indicate bin widths. }
\label{dwarf_giant_profiles}
\end{figure}

\section{Comparison with field luminosity functions}

In the field the galaxy density is orders of magnitudes lower than in
the cores of clusters like Coma. Galaxies in a dense cluster environment are likely to
follow different evolutionary paths than galaxies in low density fields.
It is therefore interesting to investigate whether this is reflected in their LF
shapes. LFs of the field population
have recently been measured by a number of surveys (e.g. Loveday et
al. 1992; Lin et
al. 1996; Lin et al. 1997; Geller et
al. 1997; Marzke et al. 1998; Lin
et al. 1999). Despite the large samples that were used to
measure the LFs, controversy on the shape remains. One of the largest
local redshift samples of galaxies selected from CCD images is provided by the Sloan
Digital Sky Survey (SDSS) (Blanton et al. 2001). The SDSS survey has
multicolour $u', g' , r', i' ,z'$
photometry, whereas most of the older surveys measure LFs only for the
$B$ and $R$ bands. We transformed the SDSS LFs to our photometric
system using the transformations given in Fukugita et
al. \shortcite{fukugita_etal}. Similarly, values for the slopes in $U$, $B$, and $r$ were
obtained by interpolating between the nearest SDSS filters. The resulting SDSS LFs are shown in
Fig.~\ref{totallfs} as dashed lines. For clarity the LFs were
renormalized by adding 2.8 to log $\phi( \rm M)$ in all bands. The faint end slopes
of the field LFs increase towards shorter wavelengths, although less
significantly than in Coma. The faint end slopes of the $r$ and $B$ band LFs of field and Coma
galaxies are very similar. Interestingly, the faint end of the Coma $U$ band LF appears to be steeper
than that of the field LF. This may indicate that dwarf galaxies
at $M_{\rm B} \sim -13$ in Coma have similar colours as dwarf galaxies in the
field.

\section{Galaxy distribution}

The best indicator of the underlying mass
distribution in our data set is the \emph{r} band, since it is least influenced by episodic star
formation. We use this band to investigate the projected
galaxy density distributions. Following Driver et al. \shortcite{driver_etal} we separate our sample
into giant and dwarf galaxies. We define giant galaxies as objects with $-23.5<M_{\rm r}<-19.5$ and
dwarf galaxies as objects with $-19.5<M_{\rm r}<-16.5$. In Fig.~\ref{coverage} we plot the
projected distributions of these two types on the sky. In panels
(a) and (b) of Fig.~\ref{dwarf_giant_profiles} we plot their foreground/background
corrected projected densities and in panel (c) their ratio as function of
distance from the cluster centre. In panels (a) and (b) we also show isothermal
profiles for comparison. The giant galaxies have a
relatively small, sharply defined, area of
high overdensity limited to the central cluster area. At a distance of
0.37~$h^{-1}_{100}$ Mpc from the cluster centre their
density abruptly drops by $\sim 70$ per cent and then decreases continuously. At distances from the cluster centre larger than
0.37~$h^{-1}_{100}$ Mpc the
dwarf-to-giant ratio (D/G) could become somewhat larger than in the core of the
cluster, but this is not significant given the large uncertainties.

\subsection{NGC 4839 group}

\begin{figure*}
\center
\includegraphics[height=8cm, width=16cm]{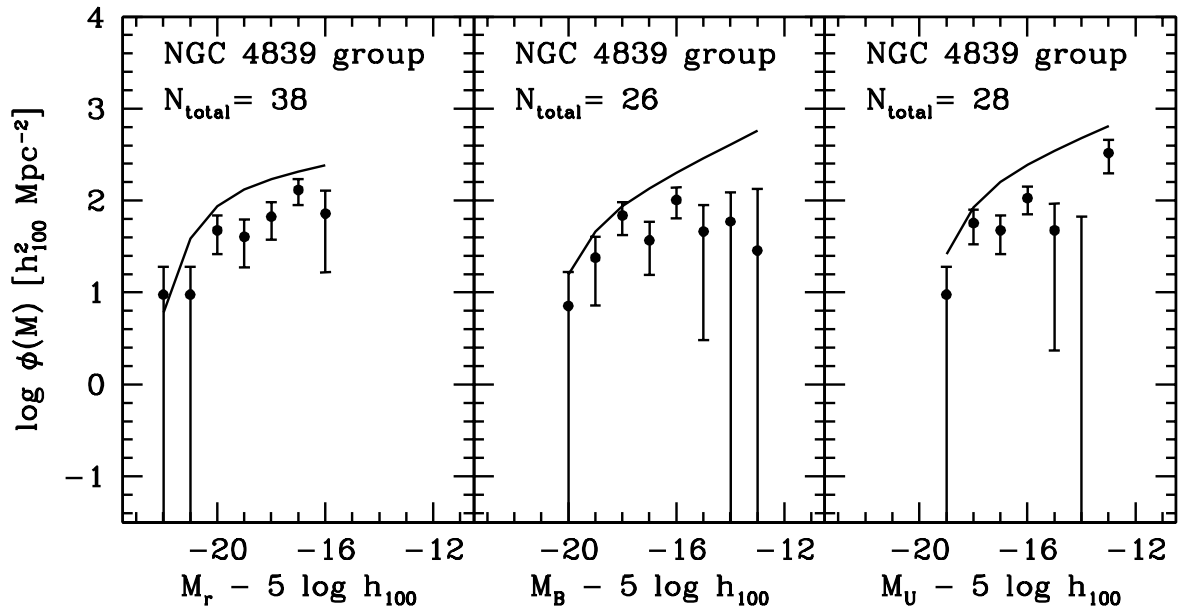}
\caption{LFs for the area around NGC 4839. The best-fitting Schechter
functions of the central luminosity functions are overplotted to
emphasize shape differences.}
\label{4839lfs}
\end{figure*}

It has long been known that at $\sim 40\arcmin$ south-west of the cluster centre a secondary concentration of galaxies exists (e.g. Wolf 1901). This is the group
associated with the cD galaxy NGC 4839. The presence of this
group is clearly visible in Fig.~\ref{dwarf_giant_profiles}: the
group's density lies well above the average density at that cluster
radius and is comparable to the central densities. We used a circular
area of 255 arcmin$^2$ centered on NGC 4839 (Fig.~\ref{circle_layout})
to determine LFs. The LFs, shown in Fig.~\ref{4839lfs}, are found to
have different shapes than the central LFs (solid lines). Mobasher \&
Trentham \shortcite{mobasher_trentham} have derived a $K$ band LF for the field
around NGC 4839. However, due to a small field size (9.9
arcmin$^2$) and large uncertainties in background subtraction, their
LF is essentially unconstrained. Lobo et al. \shortcite{lobo_etal} have derived a \emph{V} band LF with
a faint end slope comparable to their central LF, using an area of 117
arcmin$^2$. We find a $B$ band LF which is much shallower than the
central LF. The faint end of the $U$ band LF is also flatter, except
for the last point. In the $r$ band we have a lack of galaxies
fainter than $M_{\rm R}=-16$, but the faint end slope is comparable
to the slope of the central LF.

\section{Discussion}

In this paper we have used a statistical method to determine in $U$, $B$
and $r$ the LF of the galaxies in the Coma cluster and the dependence of the LF on
projected distance from the cluster centre. We did find changes in
shape as a function of radius indicating changes in the galaxy population
in the cluster. At the same time this implies that comparison with  
other work is difficult as the LF shape depends critically on the area  
chosen for study. The changes in the LF shapes are very likely related to the local
environment. We will accordingly discuss this possibility in more detail here.

The $\alpha$ parameter of the Schechter function measures effectively the faint end slope of the luminosity
distribution and characterizes the intermediate and dwarf galaxy
populations. A flat faint end slope implies a lack of these low
luminosity galaxies. L\'{o}pez-Cruz et al. \shortcite{lopez-cruz_etal}
observe the trend that steep faint end slopes are detected in poorer
clusters and the flatter slopes are, on average, found in richer
clusters. This suggests that environmental properties could dictate the
faint galaxy population. In a scenario in which mergers, interactions,
tidal stripping, destruction of dwarfs, infall etc. play a role we
expect to see a lack of faint galaxies towards the regions of higher
galaxy density. This then should be reflected in a flattening of the low luminosity end of
the LF. We do observe this effect; e.g. the slope
of the faint end of the $U$ band LF decreases from $a=0.20$ at $\sim0.7~
h_{100}^{-1}$ Mpc to $a = 0.13$ in the centre. In
terms of the slope of the Schechter function these values correspond
to $\alpha=-1.5$ and $\alpha=-1.32$. The effect is also present in the $B$ and $r$ band LFs,
and measured out to a larger radius of $r=1.3~h_{100}^{-1}$ Mpc. 

When comparing the faint end slopes of the Coma LFs with those from
the SDSS the most striking result is that the $r$ and $B$ band LFs are
very similar. This is remarkable given the differences in
environment and galaxy populations. However, the $U$ band slopes of Coma are steeper than in
the field so apparently there is a relation between the relative increase in low luminosity
systems in Coma and the colour of the band. A possible interpretation is that
the (infalling?) dwarf galaxies in the outer parts of Coma are undergoing
bursts of star formation triggered by interactions with neighbour
galaxies and/or the intra-cluster medium.
As a result the galaxies brighten, preferentially in the $U$ band, causing a steepening of the faint end slopes
of the LFs. If correct, this is a clear sign that the Coma cluster is not a
relaxed system, but that, especially in its periphery, the cluster is still
forming and inducing strong evolution to the galaxy population. The flattening of all
  Coma LFs towards the cluster centre hints that there the galaxies
  have already lost most of their gas and enhanced star formation has long
  ceased.

The group around NGC 4839 has been subject of debate in the literature, because it is not clear whether it is falling into the
cluster for the first time or has already made one pass through (for scenarios see
e.g. Colless \& Dunn 1996; White et al. 1993; Burns et
al. 1994). Bravo-Alfaro et al. \shortcite{bravo-alfaro_etal} have
done H\,{\sc i} imaging of the Coma cluster and the
NGC 4839 group. From the nondetections in the close vicinity
of NGC 4839 and the presence of several starburts and
post-starbust galaxies with very low H\,{\sc i} content in that zone
they conclude  that it passed
at least once through the core. The LF of the field around NGC
4839 has a different faint
end slope than the central LFs which could be related to the
dynamical history of this group. The observed shape differences could be explained if during the passage a large fraction of the dwarf galaxies
has been stripped from the group and redistributed throughout the cluster potential. 

To explain the dependence of the shape of the LF on projected distance from the cluster centre we would need much more information. Knowledge of the
galaxy morphologies, redshifts, the morphology-density relation
(Dressler 1980; Whitmore et al. 1983), the
type-dependent LFs, H\,{\sc i} observations etc. are necessary to
derive a complete scenario in which mergers, tidal stripping, infall
etc. play a role. For instance, the observed dips and increasing
faint end slopes could be the
combined result of Coma's morphological composition together with the
shapes of the type-dependent LFs.

\section{Summary and conclusions}

We have presented the first results of a wide field photometric survey of the Coma
cluster in the \emph{U}, \emph{B} and \emph{r} bands. The derivation
of the source catalogue, along with the steps concerning the pipeline
reduction, have been discussed. Our high quality data provides a valuable
low $z$ comparison sample for studies of galaxy morphology, colour and
luminosity at higher $z$.

In this paper we have used the data to study the dependence of the galaxy LF on passband and projected distance from the cluster centre. The LFs of the complete data set cannot be represented by single Schechter functions. The central \emph{U},
\emph{B} and \emph{r} band LFs can be represented by Schechter functions
with parameters as listed in Table~\ref{centralparameters}. 
The expectation that the shape of the LF depends on environment is
confirmed. The LFs as a function of distance from the cluster centre have (very) different shapes than the central LFs and, therefore, no universal general LF exists. The faint
ends of the LFs become steeper towards the outskirts of the cluster. A
steepening of the faint ends of the LFs towards less dense regions
clearly supports the existence of environmental effects. The difference in faint end slopes of the overall LFs and those
of the field is colour dependent (strongest in $U$) and can be attributed
to enhanced star formation in the dwarf galaxy population in the outer
parts of Coma where evolution apparently is still very strong. The LFs of the field around NGC
4839 support
the idea that this group has passed through the cluster centre at
least once. 

 We are in
the process of creating a catalogue of all spectroscopically confirmed
cluster members. This will enable us to revise the Coma LFs and to check the quality of the statistical method to determine LFs. We have also obtained Westerbork Synthesis Radio Telescope (WSRT) H\,{\sc i} mosaic data covering an total area of
$2.94\degr\times2.05\degr$ in order to study the H\,{\sc i} properties
as function of environment and assess the importance of merging and
stripping. Combined with our optical data set this will greatly
enhance the ability to study the structure and dynamics of Coma, using
a data set that is unrivalled by what is available for any other cluster.

\section*{Acknowledgements}

We thank the referee for detailed and constructive comments. Part of the data was made publically available through the Isaac Newton
Groups' Wide Field Camera Survey Programme. P. G. v. D. acknowledges support by NASA through Hubble Fellowship
grant HF-01126.01-99A awarded by the Space Telescope Science
Institute, which is operated by the Association of Universities for
Research in Astronomy, Inc., for NASA under contract NAS 5-26555.

\end{document}